# Diffusion probabilistic LMS algorithm

Sihai Guan[1], Chun Meng[1], Bharat Biswal[1,2]

**Abstract** In this paper, a novel diffusion estimation algorithm is proposed from a probabilistic perspective by combining diffusion strategy and the probabilistic least-mean-squares (PLMS) at all agents. The proposed method diffusion probabilistic LMS (DPLMS) is more robust to input signal and impulsive interference than the DSE-LMS, DRVSSLMS and DLLAD algorithms. Instead of minimizing the estimate error, the DPLMS algorithm is derived from approximating the posterior distribution with an isotropic Gaussian distribution. The stability of mean performance and computational complexity are analyzed theoretically. Results from the simulation indicate that the DPLMS algorithm is more robust to input signal and impulsive interference than the DSE-LMS, DRVSSLMS and DLLAD algorithms. These results suggest that the DPLMS algorithm can perform better in identifying the unknown coefficients under the complex and changeable impulsive interference environments.

## 1 Introduction

In recent years, distributed estimation has received more attention in multi-task networks and single-task networks [1]-[6]. In a single-task network, all nodes estimate a common parameter vector collaboratively, while in a multi-task mode each node estimates its own parameter vector [1]. Since distributed processing techniques rely on the local cooperation and data processing, three cooperation strategies for distributed estimation over networks have been widely used, i.e. the incremental, consensus, and diffusion strategies [1][5]. However, the incremental solution suffers from a number of drawbacks for real-time adaptation and learning over networks, while the consensus technique suffers from the asymmetry problem which can cause an unstable growth in the state of the network. Interestingly enough, the diffusion strategies is more adaptive to the environmental changes and can remove this asymmetry problem. Since then a variety of distributed estimation algorithms have been proposed [7]-[18].

When computing distributed estimation, an important challenge is measuring noise, since, it will significantly affect the accuracy of estimation, and most of the distributed estimation algorithms suffer from impulsive interference. It is therefore necessary to design robust distributed algorithm for impulse noises of different

Corresponding author: ✉ Sihai Guan (gcihey@sina.cn); ✉ Bharat Biswal (bbiswal@gmail.com); ✉ Chun MENG (chunmeng@uestc.edu.cn)

[1] School of Life science and technology, University of Electronic Science and Technology of China, No. 2006, Xiyuan Ave, West Hi-Tech Zone, Chengdu, Sichuan, 611731, China.

[2] Department of Biomedical Engineering, New Jersey Institute of Technology (NJIT), Newark, NJ, USA.

intensities. Recently, the diffusion least-mean *p*-power (DLMP) algorithm with fixed power *p* for the distributed parameter estimation in *α*-stable noise environments has been proposed [11]. For a given signal model, the performance of the LMP type methods is affected by *p*. In addition, to improve the robustness of distributed estimation against impulsive interference, a diffusion sign subband adaptive filtering (DSSAF) algorithm was presented [12]. However, if the unknown vector has a small number of entries, the computational complexity of the DSSAF algorithm may increase. Besides, Ni and colleagues [13] presented a diffusion sign-error LMS (DSE-LMS) algorithm, which was obtained by modifying the DLMS algorithm [2] and then applying the sign operation to the error signals. Based on reference [13], Ye and colleagues provided a steady-state and stability analyses of DSE-LMS algorithm [19]. The DSE-LMS algorithm is simple and can be implemented easily. But a lower stable-state error is the main defect of the DSE-LMS algorithm. Based on the Huber objective function, a similar set of algorithms by Wei and colleagues [16], Zhi [15] and Soheila Ashkezari-Toussi and colleagues [18] have been proposed as the DRVSS-LMS [16], DNHuber [15] and RDLMS [18] algorithms, respectively. But the RDLMS algorithm is not designed for impulse noise. For the DNHuber algorithm, the discussion of impulsive interference and input signal can be more comprehensive. Besides, the DRVSS-LMS algorithm has high algorithm complexity, which is not conducive to the implementation of practical engineering. In addition, based on least logarithmic absolute difference cost function is resistant to impulsive interference, Feng Chen and colleagues proposed the DLLAD algorithm [17]. In it, analysis of the robustness of this algorithm to input signal and impulse noise has not been performed.

However, in practical engineering applications, there will be impulsive interference with various characteristics so that the input signal may not behave to be as ideal as assumed, for example, the input signal is going to be strongly correlated. Thus, from approximating the posterior distribution with an isotropic Gaussian distribution, we proposed a diffusion probabilistic least-mean-squares (DPLMS) algorithm which is combining the ATC diffusion strategy and the PLMS algorithm [20][21] at all agents. The stability of mean performance and computational complexity are analyzed theoretically. Simulation results indicate that the DPLMS algorithm is more robust to input signal and impulsive interference than the DSE-LMS, DRVSSLMS and DLLAD algorithms.

The following parts of this paper is organized as follows: The proposed DPLMS algorithm is described in detail in Section 2. The statistical behavior of the DPLMS algorithm, including the mean performance, computation complexity, is studied in Section 3. The simulation experiments are reported in Section 4. Finally, the conclusion is provided in Section 5.

## 2 Proposed the DPLMS Algorithm

### 2.1 Modify of the PLMS algorithm

The probabilistic LMS is a VSS-LMS algorithm [21]. It has been derived from a probabilistic perspective, such as maximum a posteriori (MAP) estimation. Now we will review and modify of the PLMS algorithm in this subsection. Firstly, Consider an unknown time-varying system with length $M$, its coefficients at time $i$,
$$\mathbf{W}_o(i) = \mathbf{W}_o(i-1) + \boldsymbol{p}(i) \tag{1}$$
In Eq. (1), $\boldsymbol{p}(i)$ is a white Gaussian noise with zero-mean. The time-varying $\mathbf{W}_o(i)$ obeys the following Gaussian-distribution:
$$p(\mathbf{W}_o(i)|\mathbf{W}_o(i-1)) = N(\mathbf{W}_o(i); \mathbf{W}_o(i-1), \boldsymbol{I}\sigma_p^2) \tag{2}$$
where $\boldsymbol{I}\sigma_p^2 = \mathrm{E}[\boldsymbol{p}(i)\boldsymbol{p}^\mathrm{T}(i)]$, $\sigma_p^2$ is variance with respect to $\boldsymbol{p}(i)$, and $\boldsymbol{I}$ is an matrix with appropriate size.

For this unknown time-varying system, when we input a signal $\mathbf{X}(i) = [x(i), x(i+1), x(i+2), \cdots, x(i+M-1)]^T$, the desired signal is $d(i)$,
$$d(i) = \mathbf{W}_o^\mathrm{T}(i)\mathbf{X}(i) + \varepsilon(i) \tag{3}$$
where $\varepsilon(i)$ is a stationary additive noise with zero mean and variance of $\sigma_\varepsilon^2$. In addition. We assume that Eq.(3) obeys the following Gaussian-distribution also,
$$p(d(i)|\mathbf{X}(i), \mathbf{W}_o(i)) = N(d(i); \mathbf{W}_o^\mathrm{T}(i)\mathbf{X}(i), \sigma_\varepsilon^2) \tag{4}$$

Supposing that the estimate mean vector and variance of $\mathbf{W}_o(i)$ at iteration $i$ are $\theta(i)$ and $\sigma^2(i)$, respectively. So, by using an isotropic spherical Gaussian distribution (as Eq. (5)) to approximate the posterior distribution, we can obtained the probabilistic LMS (PLMS) algorithm.
$$p(\mathbf{W}_o(i)|\mathbf{Z}_i) = N(\mathbf{W}_o(i); \theta(i), \boldsymbol{I}\sigma^2(i)) \tag{5}$$
where $\mathbf{Z}_i = \{\mathbf{X}(k), d(k)\}_{k=1}^i$.
So, based on Eq. (5), we can get
$$p(\mathbf{W}_o(i)|\mathbf{Z}_{i-1}) = \int p(\mathbf{W}_o(i)|\mathbf{W}_o(i-1))p(\mathbf{W}_o(i-1)|\mathbf{Z}_{i-1})d\mathbf{W}_o(i-1)$$
$$= N(\mathbf{W}_o(i); \theta(i-1), \boldsymbol{I}(\sigma^2(i-1) - \sigma_\rho^2)) \tag{6}$$
where $p(\mathbf{W}_o(i-1)|\mathbf{Z}_{i-1}) = N(\mathbf{W}_o(i-1); \theta(i-1), \boldsymbol{I}\sigma^2(i-1))$ at iteration $i-1$. Combine Bayes' rule and Eq. (6), then we approximate the posterior with an isotropic Gaussian as Eq. (7).
$$p(\mathbf{W}_o(i)|\mathbf{Z}_i) = p(d(i)|\mathbf{u}(i), \mathbf{W}_o(i))p(\mathbf{W}_o(i)|\mathbf{Z}_{i-1}) = N(\mathbf{W}_o(i); \theta(i), \boldsymbol{I}\sigma^2(i)) \tag{7}$$
where
$$\theta(i) = \theta(i-1) + \alpha(i)[d(i) - \mathbf{X}^\mathrm{T}(i)\theta(i-1)]\mathbf{X}(i) \tag{8}$$
and
$$\sigma^2(i) = \left[1 - \frac{\frac{\sigma^2(i-1) + \sigma_\rho^2}{[\sigma^2(i-1) - \sigma_\rho^2]\|\mathbf{X}(i)\|^2 + \sigma_\varepsilon^2}\|\mathbf{X}(i)\|^2}{L}\right][\sigma^2(i-1) - \sigma_\rho^2] \tag{9}$$

Set $\alpha(i) = \frac{\sigma^2(i-1) + \sigma_\rho^2}{[\sigma^2(i-1) - \sigma_\rho^2]\|\mathbf{X}(i)\|^2 + \sigma_\varepsilon^2}$, then $\sigma^2(i) = \left[1 - \frac{\alpha(i)\|\mathbf{X}(i)\|^2}{L}\right][\sigma^2(i-1) - \sigma_\rho^2]$.

So, based on Eq. (7), by using MAP, i.e., $\mathbf{W}(i) = argmax_{\mathbf{W}_o(i)} p(\mathbf{W}_o(i)|\mathbf{Z}_i)$, we can get the recursive estimation weight-vector equation of the PLMS algorithm.

$$\mathbf{W}(i+1) = \mathbf{W}(i) + \alpha(i)e(i)\mathbf{X}(i) \tag{10}$$

where $e(i)$ is the error signal, as $e(i) = d(i) - y(i)$,

$$\begin{aligned} e(i) &= d(i) - \mathbf{W}^T(i-1)\mathbf{X}(i) \\ &= \mathbf{W}_o^T(i)\mathbf{X}(i) + \varepsilon(i) - \mathbf{W}^T(i-1)\mathbf{X}(i) \\ &= \mathbf{X}^T(i)(\mathbf{W}_o(i-1) + \boldsymbol{p}(i)) - \mathbf{W}^T(i-1)\mathbf{X}(i) + \varepsilon(i) \\ &= \mathbf{X}^T(i)[\mathbf{W}_o(i-1) - \mathbf{W}(i-1)] + \mathbf{X}^T(i)\boldsymbol{p}(i)) + \varepsilon(i) \\ &= \mathbf{X}^T(i)\widehat{\mathbf{W}}(i-1) + \mathbf{X}^T(i)\boldsymbol{p}(i)) + \varepsilon(i) \end{aligned} \tag{11}$$

where $\widehat{\mathbf{W}}(i-1) = \mathbf{W}_o(i-1) - \mathbf{W}(i-1)$ is the weight-deviation-vector.

To facilitate comparative analysis, we make a slight adjustment to the PLMS, that is, add a constant value parameter in Eq. (8) and Eq. (10), as Eq. (12) and Eq. (13).

$$\theta(i) = \theta(i-1) + \tau\alpha(i)[d(i) - \mathbf{X}^T(i)\theta(i-1)]\mathbf{X}(i) \tag{12}$$

$$\mathbf{W}(i+1) = \mathbf{W}(i) + \tau\alpha(i)e(i)\mathbf{X}(i) \tag{13}$$

where $0 < \tau \leq 1$ is a constant value parameter.

**2.2 The DPLMS algorithm**

We consider a network composed of $N$ nodes, where each node measures its data $\{\mathbf{X}_n(i), d_n(i)\}$ to estimate a unknown parameter vector $\mathbf{W}_o$ of length $M$ vector through a linear model at agent $n \in \{1, 2, \cdots, N\}$:

$$d_n(i) = \mathbf{X}_n(i)\mathbf{W}_o + \varepsilon_n(i) \tag{14}$$

where $\varepsilon_n(i)$ is the measurement noise with variance $\sigma_{\varepsilon,n}^2$.

*Assumption 1:* $\mathbf{X}_n(i)$ *is zero-mean Gaussian, temporally white and spatially independent with* $\mathbf{R}_{xx,n}(i) = E[\mathbf{X}_n^T(i)\mathbf{X}_n(i)] > \mathbf{0}$.[15]

*Assumption 2: All measurement noises* $\{\varepsilon_n(i)\}$ *are independent of any other signals.*

The global cost function of DPLMS can be formulated as:

$$\begin{aligned} J^{global}(\mathbf{W}(i)) &= \sum_n J_n^{local}(\mathbf{W}(i)) \\ &= \sum_n J_n^{local}\left(argmax_{\mathbf{W}_o(i)} p(\mathbf{W}_o(i)|\mathbf{Z}_i)\right) \\ &= \sum_n \sum_{l \in N_n} c_{l,n}\left(argmax_{\mathbf{W}_o(i)} p(\mathbf{W}_o(i)|\mathbf{Z}_i)\right) \end{aligned} \tag{15}$$

where the set of nodes that are connected to $n$ (including $n$ itself) is denoted by $N_n$ and is called the neighborhood of nodes $n$. The weighting coefficients $\{c_{l,n}\}$ are real and satisfy $\sum_{l \in N_n} c_{l,n} = 1$. $\{c_{l,n}\}$ forms a nonnegative combination matrix $C$.

Therefore, Inspired by DLMS in [2], we get the inspiration that we can refer to this process to design the DPLMS algorithm via two steps, adaptation and combination:
Adaptation step:

$$\boldsymbol{\varphi}_n(i) = \mathbf{W}_n(i-1) + \mu\alpha_n(i)e_n(i)\mathbf{X}_n(i) \tag{16}$$

combination step:

$$\mathbf{W}_n(i) = \sum_{l\in N_n} a_{l,n}\boldsymbol{\varphi}_l(i) \tag{17}$$

where $\mu$ is the step size (learning rate), and $\boldsymbol{\varphi}_n(i)$ is the local estimates at node $n$. The weighting coefficients $\{a_{l,n}\}$ is real, and $a_{l,n} = 0$ if $l \notin N_n$. $\{a_{l,n}\}$ forms a nonnegative combination matrix $\mathbf{A}$ with $\mathbf{A}^T\mathbf{1}=1$. Then, we combine the ATC diffusion strategy and the probabilistic least-mean-squares (PLMS) at all agents. For convenience, a summary of the procedure for the DPLMS algorithm based on the analysis presented above is given in Table 1.

Table 1 the DPLMS algorithm summary

---
Initialize: $\{w_{n,0} = 0\}$ for each node $n$, step size $\mu$, and set nonnegative combination weights $\{a_{l,n}, c_{l,n}\}$ for each time $i \geq 0$ and each agent $n$, and repeat:
**for** $i = 1: num$ber
    **for each node** $n$
    **adaptation**

$$\sigma^2{}_n(i) = \left[1 - \frac{\alpha_n(i)\|\mathbf{X}_n(i)\|^2}{L}\right]\left[\sigma^2{}_n(i-1) - \sigma^2_{\rho,n}\right]$$

$$\alpha_n(i) = \frac{\sigma_n^2(i-1) + \sigma_\rho^2}{[\sigma_n^2(i-1) - \sigma_{\rho,n}^2]\|\mathbf{X}_n(i)\|^2 + \sigma_{\varepsilon,n}^2}$$

$$\boldsymbol{\varphi}_n(i) = \mathbf{W}_n(i-1) + \mu\alpha_n(i)e_n(i)\mathbf{X}_n(i)$$

  **combination**

$$\mathbf{W}_n(i) = \sum_{l\in N_n} a_{l,n}\boldsymbol{\varphi}_l(i)$$

    $N_n$ is the neighbor nodes of node $n$ in communication subnet-work.
**end for**

---

## 3. Performance of the DPLMS algorithm

Performances of the DPLMS algorithm include mean behavior and computational complexity will be discussed in this subsection. Firstly, let us give some equations, $\widehat{\mathbf{W}}_n(i-1) = \mathbf{W}_o - \mathbf{W}_n(i-1)$, $\widehat{\boldsymbol{\varphi}}_n(i-1) = \mathbf{W}_o - \boldsymbol{\varphi}_n(i-1)$, $\mathbf{W}(i) = \text{col}\{\mathbf{W}_1(i), \mathbf{W}_2(i), \cdots, \mathbf{W}_M(i)\}$, $\boldsymbol{\varphi}(i) = \text{col}\{\boldsymbol{\varphi}_1(i), \boldsymbol{\varphi}_2(i), \cdots, \boldsymbol{\varphi}_M(i)\}$, $\widehat{\mathbf{W}}(i) = \text{col}\{\widehat{\mathbf{W}}_1(i), \widehat{\mathbf{W}}_2(i), \cdots, \widehat{\mathbf{W}}_M(i)\}$, $\widehat{\boldsymbol{\varphi}}(i) = \text{col}\{\widehat{\boldsymbol{\varphi}}_1(i), \widehat{\boldsymbol{\varphi}}_2(i), \cdots, \widehat{\boldsymbol{\varphi}}_M(i)\}$.

**3.1 Mean performance analysis**
Using the above definitions $\widehat{\mathbf{W}}_n(i-1) = \mathbf{W}_o - \mathbf{W}_n(i-1)$, Eq. (16) can be written as

$$\widehat{\boldsymbol{\varphi}}(i) = \widehat{\mathbf{W}}_n(i-1) - \mathbf{S}(i)\left[\mathbf{R}(i)\widehat{\mathbf{W}}_n(i-1) + \mathbf{g}(i)\right] \tag{18}$$

where $\mathbf{S}(i) = \text{diag}\{\mu\alpha_n(i)\mathbf{I}_M, \cdots, \mu\alpha_n(i)\mathbf{I}_M\}$, $\mathbf{R}(i) = \text{diag}\{\sum_{l=1}^N c_{l,1}\mathbf{X}_l^T(i)\mathbf{X}_l(i),$

$\cdots, \sum_{l=1}^N c_{l,N}\mathbf{X}_l^T(i)\mathbf{X}_l(i)\}$, $\mathbf{O}(i) = \mathbf{C}^T\text{col}\{\mathbf{X}_n^T(i)\varepsilon_n(i), \mathbf{X}_n^T(i)\varepsilon_n(i), \cdots, \mathbf{X}_n^T(i)\varepsilon_n(i)\}$,

$\mathbf{C} = \mathbf{C} \otimes \mathbf{I}_M$, and $\otimes$ denotes the Kronecker product operation.

Also combine $\widehat{\boldsymbol{\varphi}}_n(i-1) = \mathbf{W}_o - \boldsymbol{\varphi}_n(i-1)$ and Eq. (17), we can get
$$\widehat{\mathbf{W}}(i) = \mathbf{A}^T \widehat{\boldsymbol{\varphi}}(i) \tag{19}$$

where $\mathbf{A} = \mathbf{A} \otimes \mathbf{I}_M$.

So, taking Eq. (18) into Eq. (19), we can get Eq. (20), as
$$\widehat{\mathbf{W}}(i) = \mathbf{A}^T[\mathbf{I} - \mathbf{S}(i)\mathbf{R}(i)]\widehat{\mathbf{W}}(i-1) + \mathbf{A}^T\mathbf{S}(i)\mathbf{O}(i) \tag{20}$$

Then taking the expectation of Eq. (20),
$$\begin{aligned} \mathrm{E}[\widehat{\mathbf{W}}(i)] &= \mathrm{E}\{\mathbf{A}^T[\mathbf{I} - \mathbf{S}(i)\mathbf{R}(i)]\widehat{\mathbf{W}}(i-1) + \mathbf{A}^T\mathbf{S}(i)\mathbf{O}(i)\} \\ &= \mathbf{A}^T[\mathbf{I} - \mathrm{E}\{\mathbf{S}(i)\mathbf{R}(i)\}]\mathrm{E}\{\widehat{\mathbf{W}}(i-1)\} + \mathrm{E}\{\mathbf{A}^T\mathbf{S}(i)\mathbf{O}(i)\} \\ &= \mathbf{A}^T[\mathbf{I} - \mathrm{E}\{\mathbf{S}(i)\mathbf{R}(i)\}]\mathrm{E}\{\widehat{\mathbf{W}}(i-1)\} \end{aligned} \tag{21}$$

where $\mathrm{E}[\mathbf{S}(i)\mathbf{R}(i)] = \mu \mathrm{diag}\{\sum_{l=1}^N \alpha_l(i)c_{l,1}\mathbf{R}_{xx,l}(i), \cdots, \sum_{l=1}^N c_{l,N}\mathbf{R}_{xx,l}(i)\}$.

Condition for stability of the mean weight error vector (as Eq. 21) is given by
$$0 < \mu < \frac{2}{\rho_{max}(\sum_{l=1}^N \alpha_l(i)c_{l,l}\mathbf{R}_{xx,l}(i))} \tag{22}$$

where $\rho_{max}$ denotes the maximal eigenvalue of $\sum_{l=1}^N \alpha_l(i)c_{l,l}\mathbf{R}_{xx,l}(i)$. So, based on Eq. (22), we obtain $\mathrm{E}[\widehat{\mathbf{W}}(\infty)] = \mathbf{0}$.

### 3.2 Computational complexity

For convenience, computational complexity of the DPLMS algorithm and that of other existing algorithms is summarized in Table 2. From Table2, compared to the DSE-LMS [19], DRVSSLMS [16] algorithms, the DPLMS algorithm has smaller computational complexity. The computational complexity of DLLAD [17] and DPLMS is the same.

Table 2. Computational complexity of the DSE-LMS, DRVSSLMS, DLLAD and DPLMS algorithms

| Algorithm | Computational cost per iteration | | | | |
|---|---|---|---|---|---|
| | Recursion | × | + | \| \| | sgn( ·) |
| DSE-LMS | Eq. 1*a* in [19] | (2*M*+1)*N*+*M* | (3*M*-1)*N* | 0 | *N* |
| | Eq. 1*b* in [19] | *NM* | (*N*-1)*M* | 0 | 0 |
| DRVSSLMS | Eq. 11 in [16] | > (3*M*+1)*N* +*M* | (3*M*-1)*N* | 0 | 0 |
| | Eq. 11 in [16] | > (3*M*+1)*N*+*M* | (3*M*-1)*N* | 0 | *N* |
| | Eq. 12 in [16] | *NM* | (*N*-1)*M* | 0 | 0 |
| DLLAD | Eq. 16 in [17] | 2*MN*+*M* | (3*M*-1)*N* | *N* | 0 |
| | Eq. 17 in [17] | *NM* | (*N*-1)*M* | 0 | 0 |
| DPLMS | Eq. 16 in this paper | 2*MN*+*M* | (3*M*-1)*N* | 0 | 0 |
| | Eq. 17 in this paper | *NM* | (*N*-1)*M* | 0 | 0 |

where "×" denotes Multiplications. ">" denotes larger than. "+" denotes Additions. "| |" denotes Absolute.

## 4. Simulation results

Performance of the proposed DPLMS algorithm by comparing it with that of the DSE-LMS [19], DRVSSLMS [16] and DLLAD [17] algorithms is evaluated in this section. Several experiments were performed in a system identification application in presence of impulsive interference (impulse noise with a Bernoulli-Gaussian distribution was added to the system output.) and Gaussian noise (white Gaussian random process with zero mean and variance = 0.01). For this unknown system, we set *M*=16, and the parameters vector is selected randomly. Network topology with *N*=20 nodes. Besides, in [15], we already know that how to get the Bernoulli-Gaussian impulse noise. Set impulse noise as $v(i) = f(i)g(i)$ is a product of a Bernoulli process $g(i)$ and a Gaussian process $f(i)$, where $f(i)$ is a white Gaussian random process with zero mean and variance $\sigma_f^2$, and $g(i) = 0,1$ is a Bernoulli process with the probabilities $p(1) = \text{Pr}$ and $p(0) = 1 - \text{Pr}$. The impulsive interference is also assumed to be spatiotemporally independent distributed with power $\sigma_f^2$. Also, $\mathbf{X}_n(i)$ at node $n$ are assumed to be spatiotemporally independent zero-mean white Gaussian distributed with different covariance matrixes $\mathbf{R}_{xx,n}$. For the adaptation and combination weights, we apply the uniform rule (i.e., $a_{l,n} = c_{l,n} = 1/N_n$, where the set of nodes that are connected to *n* is denoted by $N_n$). The performance of algorithms is evaluated by the network mean square deviation (MSD) as $\text{MSD}(i) = \frac{1}{N}\sum_{n=1}^{N} \text{E}[|\mathbf{W}_o - \mathbf{W}_n(i)|^2]$. The results are obtained via Monte Carlo simulation using 60 independent run sets and an iteration number of 4000.

*Experiment 1*

In order to illustrative our algorithm is more robust to input signal and has the faster convergence rate and lowest steady-state error than the DSE-LMS, DRVSSLMS and DLLAD algorithms. We design this experiment, in this experiment, there have same network topology and same impulsive interference. Two nodes are declared neighbors probability (probability=0.2) from each other are declared neighbors, network topology is Fig. 1. The MSD curves for DSE-LMS (*μ*=0.6), DRVSSLMS (*μ*=0.6), DNLMS (*μ*=0.6), and DLLAD (*μ*=0.6) algorithms in Fig.2, Fig.3, and Fig.4 are different types of input signal when the measurement noise is impulsive interference with Pr=0.4, and $\sigma_f^2 = 0.2$.

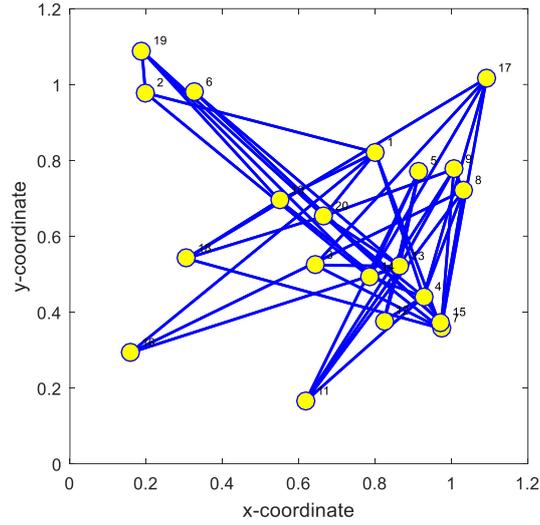

Fig.1. Random network topology to be decided by probability.

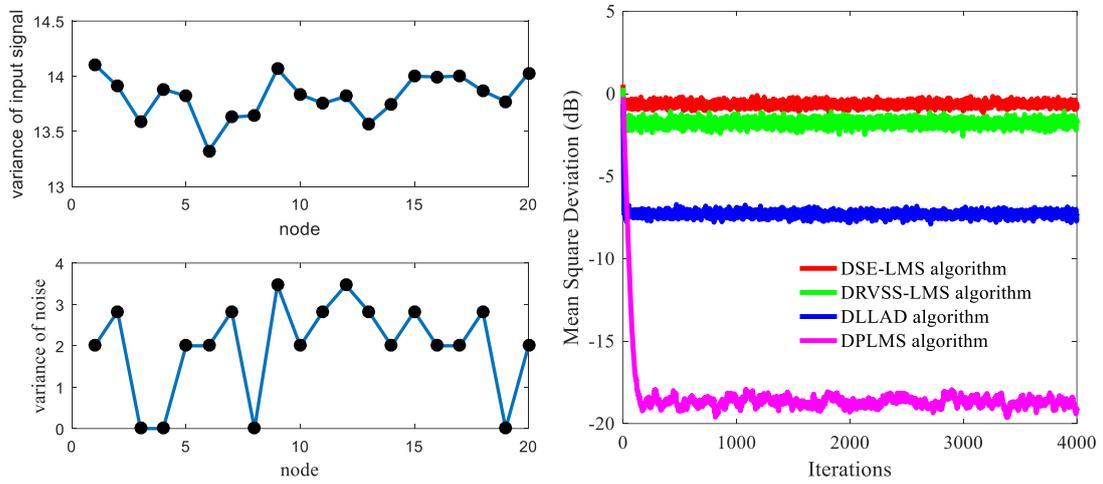

Fig.2. (Left_top) Regressor variances, (Left_bottom) Noise variances; (Right) Simulation MSD curves of the DSE-LMS, DRVSSLMS, DLLAD and DPLMS algorithms with $\mathbf{R}_{xx,n} = \sigma_{x,n}^2 \mathbf{I}_M$.

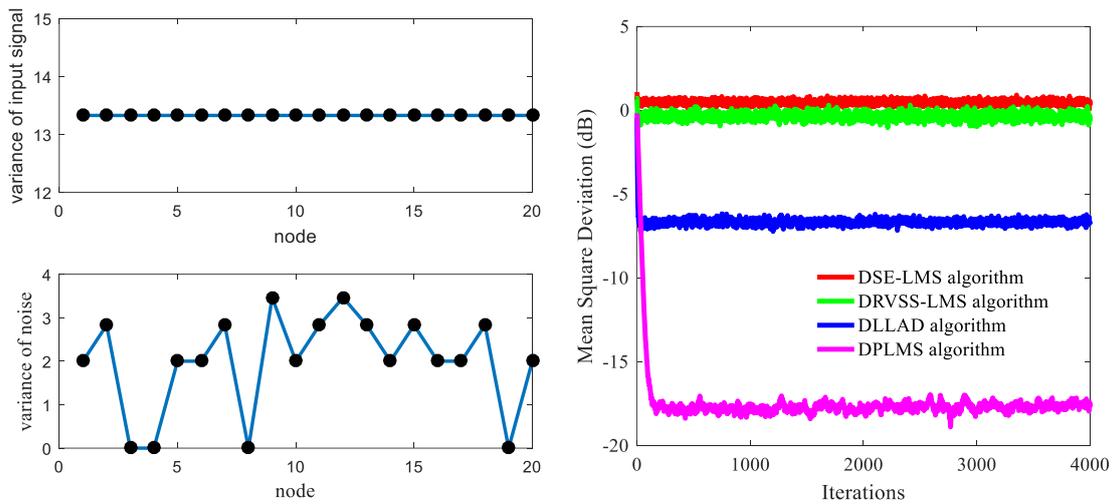

Fig.3. (Left_top) Regressor variances, (Left_bottom) Noise variances; (Right) Simulation MSD curves of the DSE-LMS, DRVSSLMS, DLLAD and DPLMS algorithms with $\mathbf{R}_{xx,n} = \sigma_{x,n}^2 \mathbf{I}_M$.

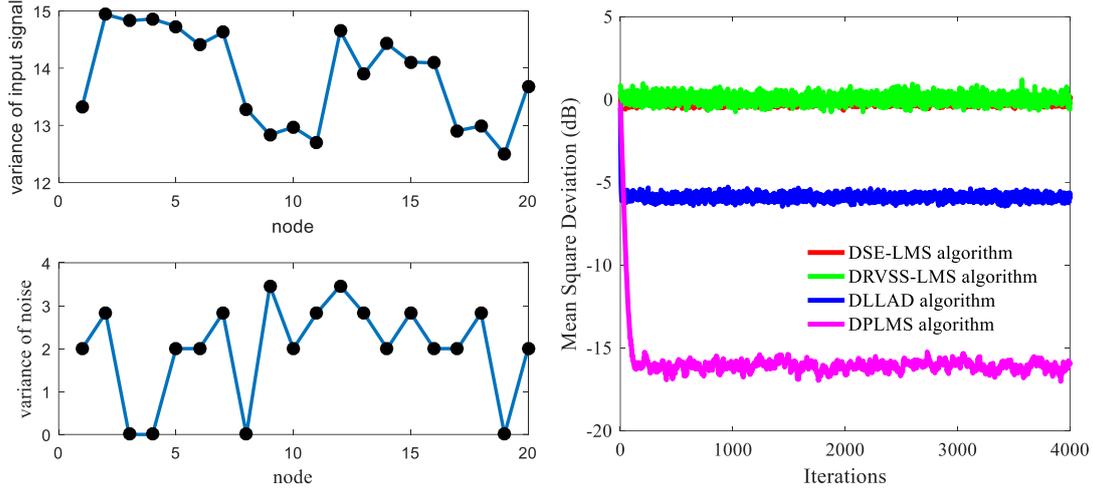

Fig.4. (Left_top) Regressor variances, (Left_bottom) Noise variances; (Right) Simulation MSD curves of the DSE-LMS, DRVSSLMS, DLLAD and DPLMS algorithms with $\mathbf{R}_{xx,n} = \sigma_{x,n}^2(i)\mathbf{I}_M, i = 1,2,3,\cdots,M$.

From Fig. 2, Fig.3 and Fig.4, under the same noise interference intensity and network structure, although the input signals for different characteristics are used, the DPLMS algorithm has the faster convergence rate and lowest steady-state error than the DSE-LMS, DRVSSLMS and DLLAD algorithms. Besides, the DPLMS algorithm is more robust to input signal. In conclusion, from ***Experiment 1***, we can get the DPLMS algorithm is obviously superior to the DSE-LMS, DRVSSLMS and DLLAD algorithms.

*Experiment 2*

In order to illustrative our algorithm is more robust to impulsive interference and has the faster convergence rate and lowest steady-state error than the DSE-LMS, DRVSSLMS and DLLAD algorithms. We design this experiment, in this experiment, there have same network topology and same input signal. Two nodes that are within a certain radius (radius=0.3) to determine which nodes are neighbors, network topology is Fig. 5(Left) and $\mathbf{R}_{xx,n}$ is a diagonal matrix with possibly different diagonal entries chosen randomly as Fig. 5(Right). The MSD curves for DSE-LMS ($\mu$=0.4), DRVSSLMS ($\mu$=0.4), DNLMS ($\mu$=0.4), and DLLAD ($\mu$=0.4) algorithms in Fig.6 with Pr=0.1, Pr=0.4, and Pr=0.7 under same $\sigma_f^2 = 0.2$, respectively.

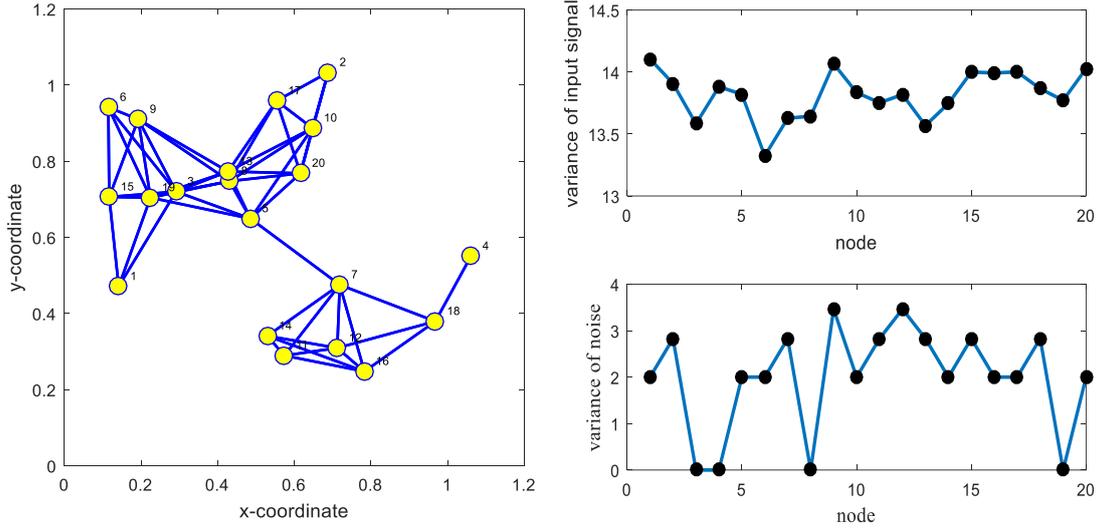

Fig.5. (Left) Random network topology to be decided by distance; (Right_top) Regressor variances, (Right_bottom) Noise variances.

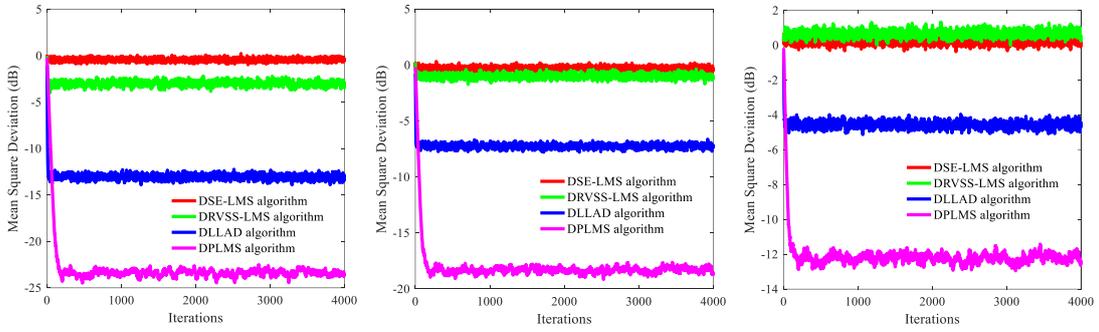

Fig.6. Simulation MSD curves of the DSE-LMS, DRVSSLMS, DLLAD and DPLMS algorithms. (Left) with Pr=0.1, (middle) with Pr=0.4 and (Right) with Pr=0.7.

From Fig. 6, under the same noise interference intensity, input signal (correlated) and network structure, although different the probability density of impulsive interference are considered, the convergence rate of the DPLMS algorithm is a slight faster than that of the DSE-LMS, DRVSSLMS and DLLAD algorithms, and he DPLMS algorithm still have a smaller steady-state misalignment than the DSE-LMS, DRVSSLMS and DLLAD algorithms. In a word, from *Experiment 2*, we can observe that the DPLMS algorithm is more robust to impulsive interference than the DSE-LMS, DRVSSLMS and DLLAD algorithms.

*Experiment 3*
In order to illustrative our algorithm is more robust to impulsive interference and has the faster convergence rate and lowest steady-state error than the DSE-LMS, DRVSSLMS and DLLAD algorithms. We design this experiment, in this experiment, there have same network topology and same input signal. Two nodes that are within a certain radius (radius=0.3) to determine which nodes are neighbors, network topology is Fig. 7. The MSD curves for DSE-LMS ($\mu$=0.4), DRVSSLMS ($\mu$=0.4), DNLMS ($\mu$=0.4), and DLLAD ($\mu$=0.4) algorithms in Fig. 8(Left-top): Pr=0.7, $\sigma_f^2 = 0.2$, Fig. 8

(Right-top): Pr=0.7, $\sigma_f^2 = 0.4$, Fig. 8 (Left-bottom): Pr=0.4, $\sigma_f^2 = 0.4$, and Fig. 8 (Right-down): Pr=0.4, $\sigma_f^2 = 0.6$, respectively.

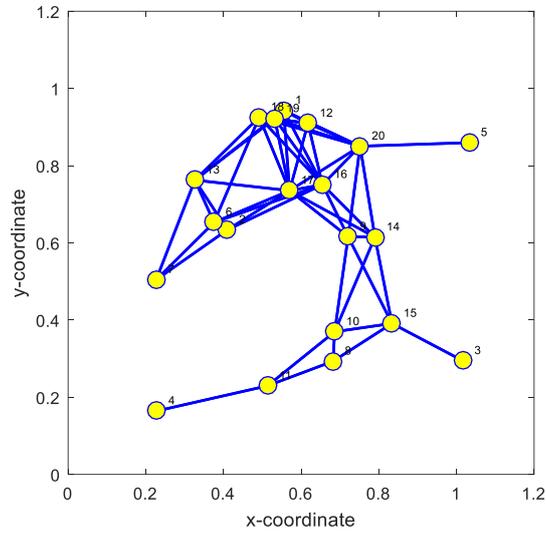

Fig.7. Random network topology to be decided by distance

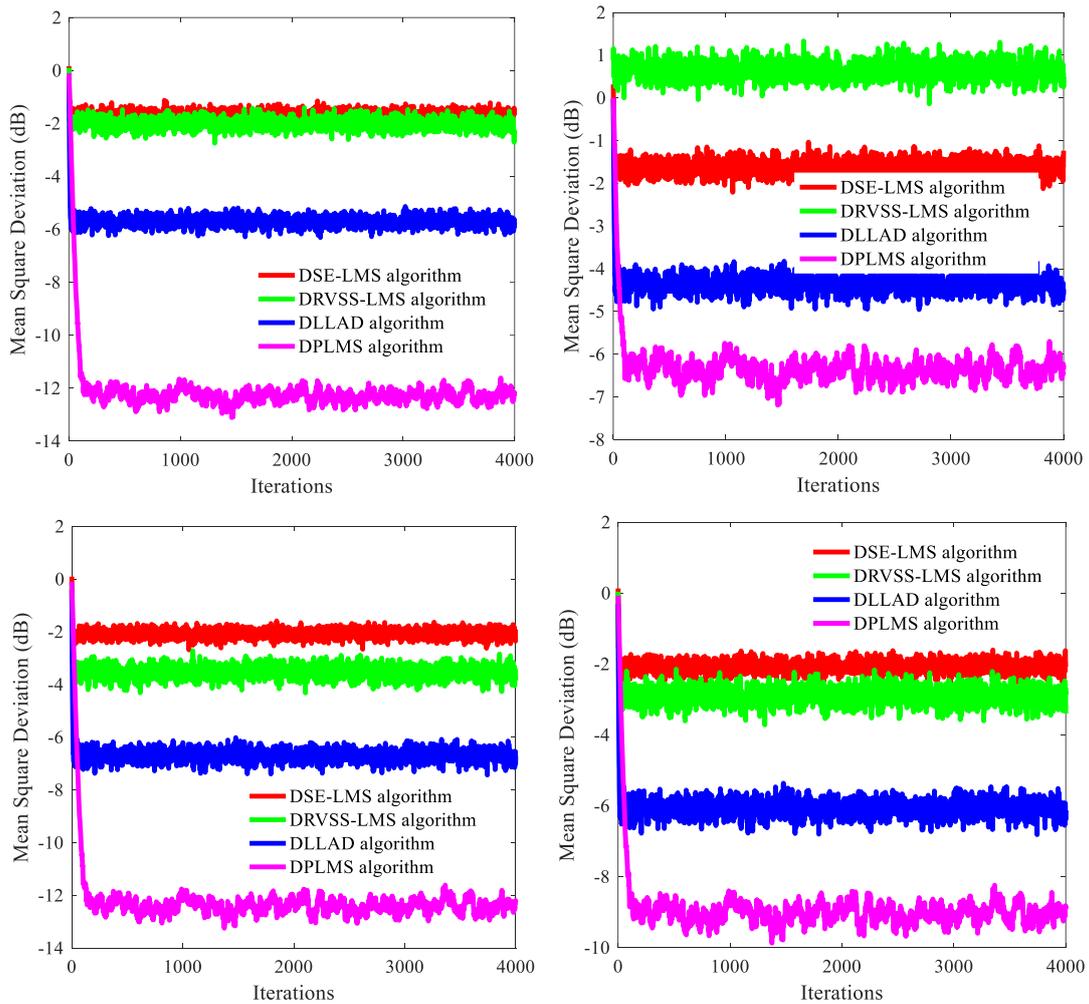

Fig.8. Simulation MSD curves of the DSE-LMS, DRVSSLMS, DLLAD and DPLMS algorithms. (Left-top): Pr=0.7, $\sigma_f^2 = 0.2$; (Right-top): Pr=0.7, $\sigma_f^2 = 0.4$; (Left-bottom): Pr=0.4, $\sigma_f^2 = 0.4$;

(Right-down): Pr=0.4, $\sigma_f^2 = 0.6$.

Comparing Fig.8 (Left_top) with Fig.8 (Right_top), the DPLMS algorithm can perform better in identifying the unknown coefficients under different impulsive interference intensity. From Fig.8 (Left_bottom) with Fig.8 (Right_bottom), the same conclusion can be obtained. However, for the DSE-LMS, DRVSSLMS and DLLAD algorithms, the identification performance is easily interfered by different characteristics, that is to say, the DPLMS algorithm is more robust. The DPLMS algorithm is obviously superior to the DSE-LMS, DRVSSLMS and DLLAD algorithms. So, for distributed estimation in impulsive interference environments, from ***Experiment 1- Experiment 3***, we know that the DPLMS algorithm can perform better in identifying the unknown coefficients under the complex and changeable impulsive interference environments.

## 5. Conclusion

This paper described a novel DPLMS algorithm, which is a distributed algorithm robust to input signal and impulsive interference environments. The method is developed based on the combination and modification of adapt-then-combine diffusion LMS (DLMS) algorithm band the PLMS algorithm at all agents to develop a DPLMS algorithm. The theoretical analysis demonstrates that DPLMS algorithm can achieve effective estimate from a probabilistic perspective. The computational complexity of the proposed algorithm is smaller than that of the DSE-LMS, DRVSSLMS algorithms, and equal to that of DLLAD algorithm. Besides, simulation results showed that the DPLMS algorithm is more robust to input signal and impulsive interference than the DSE-LMS, DRVSSLMS and DLLAD algorithms. That is to say, the DPLMS algorithm can perform better in identifying the unknown coefficients under the complex and changeable impulsive interference environments, which will have significant impact on real world applications. Besides, in this paper we wish it open the door to bring more probability theory techniques to distributed adaptive filtering algorithm.

**Acknowledgments**: This work was supported in part by the National Natural Science Foundation of China (NSFC) under Grants 61871420.

## References

[1] C.G. Lopes, A.H. Sayed, Diffusion least-mean squares over adaptive networks: formulation and performance analysis [J]. IEEE Transactions Signal Processing, 2008, 56(7):3122-3136.
[2] F.S. Cattivelli, A.H. Sayed, Diffusion LMS strategies for distributed estimation [J]. IEEE Transactions Signal Processing, 2010, 58(3):1035-1048.
[3] Tu S Y, Sayed A H. Diffusion strategies outperform consensus strategies for distributed estimation over adaptive networks [J]. IEEE Transactions on Signal Processing, 2012, 60(12):6217-6234.
[4] Lorenzo P D, Sayed A H. Sparse distributed learning based on diffusion

adaptation [J]. IEEE Transactions on Signal Processing, 2013, 61(6):1419-1433.
[5] A. H. Sayed, Adaptive networks [J], Proceedings of the IEEE, 2014, 102(4):460-497.
[6] Jie C, Richard C, Sayed A H. Diffusion LMS over multitask networks [J]. IEEE Transactions on Signal Processing, 2015, 63(11):2733-2748.
[7] Lee H S, Yim S H, Song W J. $z^2$-proportionate diffusion LMS algorithm with mean square performance analysis [J]. Signal Processing, 2017, 131:154-160.
[8] Zhang S, So H C, Mi W, et al. A family of adaptive decorrelation NLMS algorithms and its diffusion version over adaptive networks [J]. IEEE Transactions on Circuits & Systems I Regular Papers, 2017, PP (99):1-12.
[9] F.S. Cattivelli, C.G. Lopes, A.H. Sayed, Diffusion recursive least-squares for distributed estimation over adaptive networks [J]. IEEE Transactions Signal Processing, 2008, 56(5):1865-1877.
[10] Liu Y, Li C, Zhang Z. Diffusion sparse least-mean squares over networks [J]. IEEE Transactions Signal Processing, 2012, 60(8):4480-4485.
[11] Wen F. Diffusion least-mean $p$-power algorithms for distributed estimation in alpha-stable noise environments [J]. Electronics Letters. 2013, 49(21): 1355-1356.
[12] Ni J. Diffusion sign subband adaptive filtering algorithm for distributed estimation [J]. IEEE Signal Processing Letters, 2015, 22(11):2029-2033.
[13] Ni J, Chen J, Chen X. Diffusion sign-error LMS algorithm: Formulation and stochastic behavior analysis [J]. Signal Processing, 2016, 128:142-149.
[14] Wen P, Zhang J. Variable step-size diffusion normalized sign-error algorithm [J]. Circuits Systems and Signal Processing, 2018, 37(20):4993-5004.
[15] Zhi L, Guan S. Diffusion normalized Huber adaptive filtering algorithm [J]. Journal of the Franklin Institute, 2018, 355(8):3812-3825.
[16] Huang W, Li L, Li Q, et al. Diffusion robust variable step-size LMS algorithm over distributed networks [J]. IEEE Access, 2018, 6:47511-47520.
[17] Feng Chen, Tao Shi, et al. Diffusion least logarithmic absolute difference algorithm for distributed estimation [J]. Signal Processing, 2018, 142:423-430.
[18] Soheila Ashkezari-Toussi, Hadi Sadoghi-Yazdi. Robust diffusion LMS over adaptive networks [J]. Signal Processing, 2019, 158:201-209.
[19] Gao Y, Ni J, Chen J, et al. Steady-state and stability analyses of diffusion sign-error LMS algorithm [J]. Signal Processing, 2018, 149:62-67.
[20] J. Fernandez-Bes, V. Elvira, et al. A probabilistic least-mean-squares filter [C]. IEEE International Conference on Acoustics, Speech and Signal Processing, Brisbane, QLD, Australia, Aug. 2015, pp.2199-2203.
[21] Fuyi Huang, Jiashu Zhang, Sheng Zhang. Mean-square-deviation analysis of probabilistic LMS algorithm [J]. Digital Signal Processing, 2019, 92:26-35.